\documentclass[runningheads,a4paper]{llncs}

\usepackage{amssymb}
\usepackage{graphicx}
\usepackage{color}
\usepackage{times}
\usepackage{fancyhdr}
\usepackage{amssymb,amsmath}
\usepackage{lastpage}
\usepackage{epsfig}
\usepackage{algorithm2e}
\usepackage{cite}

\newcommand{\keywords}[1]{\par\addvspace\baselineskip
\noindent\keywordname\enspace\ignorespaces#1}
\begin{document}

\title{When Both Transmitting and Receiving Energies Matter: An Application of Network Coding in
Wireless Body Area Networks}

\titlerunning{An Application of Network Coding in Wireless Body Area Networks}

\author{Xiaomeng Shi$^\ast$ \and Muriel M\'edard$^\ast$ \and Daniel E. Lucani$^\dagger$}
\institute{$^\ast$Massachusetts Institute of Technology, Cambridge, MA 02139, USA\\
$^\dagger$Instituto de Telecomunica\c c\~oes, DEEC Faculdade de Engenharia, Universidade do Porto,
Portugal
\\ \{xshi, medard\}@mit.edu, dlucani@fe.up.pt}
\authorrunning{Xiaomeng Shi, Muriel M\'edard, Daniel E. Lucani}
\maketitle

\begin{abstract}
A network coding scheme for practical implementations of wireless body area networks is presented, with
the objective of providing reliability under low-energy constraints. We propose a simple network layer
protocol for star networks, adapting redundancy based on both transmission and reception energies for
data and control packets, as well as channel conditions. Our numerical results show that even for small
networks, the amount of energy reduction achievable can range from $29\%$ to $87\%$, as the receiving
energy per control packet increases from equal to much larger than the transmitting energy per data
packet. The achievable gains increase as a) more nodes are added to the network, and/or b) the channels
seen by different sensor nodes become more asymmetric.

\end{abstract}


\section{Introduction}\label{sect:introduction}
Body Area Networks (BAN) present numerous application opportunities in areas where measured personal
information is to be stored and shared with another individual or a central database. One example is
wearable medical monitors which can relay patients' vital information to physicians or paramedics in
real time. A wireless body area network (WBAN) is composed of sensors attached to the human body. The
sensors also function as transceivers to relay measurements to a personal server (base station); this
central receiver then communicates with remote servers or databases. In this paper we consider such
WBANs where the central communication problem is to ensure reliable and secure transmission of the
measured data to the base station (BS) in a timely and robust fashion. Here the amount of data uploaded
from the sensors to the BS much outweighs the amount of control signals downloaded from the BS, but
energy used to receive control signals can still be high depending on the specific physical layer
implementation and the network layer control protocol used. We develop a protocol to incorporate network
coding into the system architecture, and show for a star network with multiple sensor nodes, using
network coding can reduce the number of times sensors wake up to receive control signals, thus reducing
the overall energy consumption and lengthen the system depletion time.

In the remaining part of the introduction, we show through a simple example the potential energy gains
of applying network coding to transmissions in a WBAN. We also discuss briefly past works related to
energy efficient WBAN design. The rest of this paper is organized as follows. Section~\ref{sec:problemsetup}
describes the network and energy model, and the network coded algorithm. Section~\ref{sec:analysis}
provides a Markov chain model to analyze the optimal number of packets to transmit by each sensor node.
Numerical results are presented in Section~\ref{sec:simulations}, comparing the network coded scheme
with uncoded scheme in terms of completion energy. Section~\ref{sec:conclusions} concludes the paper.
\vspace{-4pt}

\subsection{Example: Network Coding Benefits in WBAN}\label{sec:example}
\vspace{-2pt}

\begin{figure}[!t]
\centering
\includegraphics[width=.95\textwidth,keepaspectratio,clip=true]{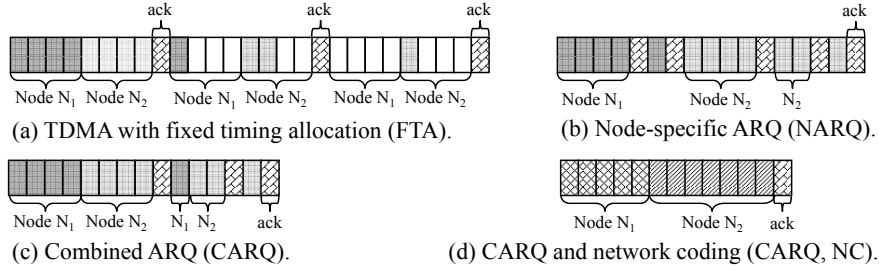}
\vspace{-8pt} \caption{Example comparing overall completion energy for $2$ nodes, each with $4$ packets
to upload; the erasure probabilities are $p_1=0.2$, $p_2=0.4$.} \label{fig:example}\vspace{-12pt}
\end{figure}

Before discussing other past work related to energy efficient WBAN design, we first show through a
simple example why network coding can be beneficial. One category of energy use often overlooked in
wireless networks is the reception energy spent on listening to control signals from the base station.
In a WBAN, however, such reception energies can have more significant effects on node depletion time
since data rate is much lower, but control signals need to be transmitted frequently for medium access
purposes. Let us consider a two-sensor star network with nodes $N_1$ and $N_2$, each trying to directly
upload $4$ packets to a BS through the same frequency band. In the link layer, assume the packet erasure
probabilities are time invariant, at $0.2$ and $0.4$ respectively. Figure~\ref{fig:example} shows
instances of four different possible communication schemes, all based on time division multiple access
(TDMA) with automatic repeat requests (ARQ). Shaded blocks represent data packets in transmission and
ack packets in reception; white blocks represent time during which nodes are idle. Some packets are lost
during transmission according to the different erasure probabilities. Define a transmission round to be
the transmission of data packets by one or more sensor nodes, followed by a broadcasted ack packet. Both
nodes wake up at the end of a transmission round to listen to the ack, which contains retransmission
requests and schedules for the next round.


\begin{table}[!t]
\centering \caption{Comparison of completion energy per accepted data packet; there are $2$ nodes in the
star each with $4$ packets to upload. $E_{TX}=$ total transmission energy; $E_{A}=$ energy spent on
listening to acknowledgement packets; $E_{tot}=$ total completion energy per accepted data packet, $\eta=$
throughput.}\vspace{-8pt} \label{tbl:example}
\begin{tabular}{|l|c|c|c|c|c|}
  \hline
                  & $E_{TX}$ & $E_{A}$ & $E_{tot}$ & $\eta$\\\hline
  (a) FTA         & $12E$    & $6E$    & $9E/4$    & $8/27$\\ \hline
  (b) NARQ        & $12E$    & $10E$   & $11E/4$   & $8/17$ \\ \hline
  (c) CARQ        & $12E$    & $6E$    & $9E/4$    & $8/15$ \\ \hline
  (d) CARQ-NC     & $12E$    & $2E$    & $7E/4$    & $8/13$ \\  \hline
\end{tabular}\vspace{-8pt}
\end{table}
\vspace{-4pt}
\begin{enumerate}
\item[(a)]\emph{Fixed Timing Allocation (FTA)}: each node is allocated $4$ slots per round, and both
wake up at the end of each round to receive the broadcasted ack.

\item[(b)]\emph{Node-specific ARQ (NARQ)}: each node transmits until all of its packets are received
successfully. The ack packet contains retransmission requests for the actively transmitting node and
scheduling information for both nodes.

\item[(c)]\emph{Combined ARQ (CARQ)}: both nodes are allocated specific transmission periods each
round, with a combined ack packet broadcasted at the end.

\item[(d)]\emph{Combined ARQ and network coding (CARQ-NC)}: each node linearly combines its $4$ data packets
before transmission. Since each coded packet represents an additional degree of freedom (dof) rather
than a distinct data packet, more than $4$ coded packets can be sent to compensate for anticipated
losses. 
\end{enumerate}
\vspace{-4pt}

To evaluate energy use, assume every data packet transmission and every ack packet reception consumes an
equal amount of $E$ units of energy. Table~\ref{tbl:example} compares the total energy required for the
schemes shown in Figure~\ref{fig:example}. Also shown is the throughput of each scheme, defined as the
total number of accepted data packets divided by the total transmission time in units of packet slots.
Excluding ack periods and time during which nodes are sleeping, all schemes require $12E$ in data
transmission. On the other hand, the energy used for ack reception varies significantly across the
different schemes. CARQ-NC (hereafter referred to as `NC') is the most energy efficient. FTA requires
less or the same amount of total energy than NARQ and CARQ, but is throughput inefficient. As the number
of nodes in the network increases, this inefficiency will become increasingly severe. CARQ outperforms
NARQ, and NC introduces further gains. It is not necessarily true that NC always transmits the same
total number of data packets as CARQ. In fact, NC sends \emph{more} packets than the required number of
dofs. Nonetheless, the added transmission energy is offset by reduced reception energy to give a smaller
overall completion energy. This specific example is extremely simple, but very similar results can be
expected as more sensor nodes are added. In the remaining parts of this paper, we will consider only the
CARQ and the NC scheme. Our goal is to determine analytically the optimal network coding and
transmission scheme such that the expected completion energy for the overall transmission is minimized.
\vspace{-4pt}

\subsection{Related Work}\label{sec:relatedWork}
\vspace{-2pt}

To make WBANs practical, one approach is to modify existing wireless sensor networks (WSN) to
suite the need of WBAN systems. References \cite{cao2009enabling, chenbody} compare WBAN with
traditional WSNs and give comprehensive overviews of recent research efforts in the design of WBAN
systems, particularly in terms of sensor devices, physical layer schemes and data link layer protocols.
In WSNs, energy is often wasted in medium access collisions, idle listening, and protocol overheads when
the desired data rate is low. Low power MAC protocols such as T-MAC, S-MAC and Wise-MAC have therefore
been proposed to introduce various degrees of synchronization into the transmission schedule, but these
schemes are not throughput efficient. Recently a WBAN specific MAC protocol has been proposed to adjust
parameters of the IEEE 802.15.4 parameter in an adaptive fashion to achieve energy efficiency
\cite{li2006ultra}, but direct modification of this existing standard also introduces redundant
communication modes. Unlike WSNs, a WBAN contains only a limited number of nodes, all positioned close to
the BS. A single-hop master-slave architecture with TDMA suffices to remove much of the energy wastage
seen in a WSN. Reference \cite{omeni2008energy} implements such an architecture, with adjustable wakeup
fallback times to mitigate possible slot overlaps. We use a similar TDMA setup in the current paper.
What we aim to achieve is to introduce network coding into the system architecture, such that overall
data transmission completion energy can be reduced by reducing the number of times sensor nodes wake up
to listen to control signals. Recent works by Lucani et al. consider the use of network coding in time
division duplex systems to minimize packet completion time, energy use, and queue length in unicast and
broadcast settings \cite{lucani2009random,lucani09icc,lucani2009broadcasting}. The current study extends
the unicast TDD case to that of a simple star network, with single-hop links between the BS and the
sensors.

\vspace{-4pt}

\section{Linear Network Coding for Energy Efficiency}\label{sec:problemsetup}
\vspace{-4pt}

\subsection{System Model}
\vspace{-2pt}

\begin{figure}[t!]
\centering
\includegraphics[width=.8\textwidth,keepaspectratio,clip=true]{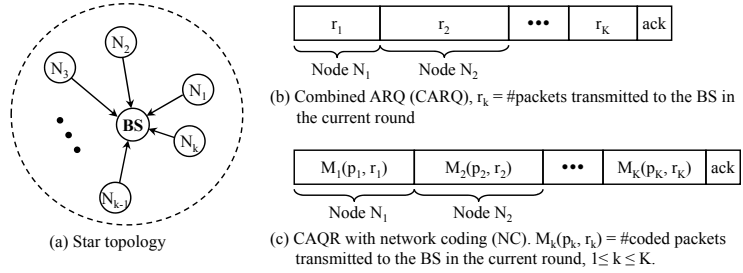}
\vspace{-6pt} \caption{Uplink transmissions using combined ARQ and network coding.}
\label{fig:scheme}\vspace{-12pt}
\end{figure}

We model a WBAN with a star topology as shown in Figure~\ref{fig:scheme}(a): each of $K$ sensor nodes
communicates with the BS directly to upload $M$ data packets. Nodes and the BS are assumed to operate in
half-duplex mode, either transmitting or receiving, but not at the same time. A WBAN occupies a single
frequency band, with the BS centrally coordinating a TDMA scheme. Exact synchronization among the nodes
and the BS is assumed, and nodes return to sleep when not transceiving. Computation of the transmission
schedule is relegated to the BS, with start and stop times allocated through the ack signal. We
assume ack packets are transmitted reliably, and propagation delays from BS to nodes are negligible. The
channel between an individual sensor $N_k$, $1\leq k \leq K$, and the BS are assumed to be memoryless,
with packet erasure probability $p_k$, which is invariant during the time when the $M$ packets are
uploaded.

The above system model may seem over-simplified, but is sufficiently accurate for the current study. As
already discussed in Section~\ref{sec:relatedWork}, the small physical size of a WBAN enables the use of
a star topology with TDMA scheduling controlled centrally by a BS. Compared to sensor nodes, the BS is
relatively unconstrained in power. Ack packets can therefore be piggybacked on a periodic
synchronization signal transmitted at high power, or protected through error correction codes to ensure
reliability. In an actual implementation, additional headers or beacon periods will be needed for
synchronization, but such details can be safely omitted here in analyzing the data transmission energy
efficiency and system throughput. An additional difficulty in WBAN design is channel modeling for
physical layer designs. Unlike cellular networks or WSNs, a WBAN is in close proximity to the human
body. Absorption of emitted power and body movement can easily and frequently alter the channel
response. Reference~\cite{yazdandoost15channel} provides a summary of channel modeling studies conducted
and submitted to the IEEE 802.15.6 body area network task group. In the current paper, we only consider
an erasure channel abstraction for the network layer model. The time-invariance assumption is a
reasonable first step, since data in WBAN come in very small bursts periodically and the channel can be
assumed to fade slowly over each such small periods.

In the CARQ scheme, nodes take turns to transmit data packets before waiting for a combined ack, which
contains repeat requests and scheduling information. Figure~\ref{fig:scheme}(b) illustrates one round
of transmission, where $r_k$ represents the number of packets requested by the BS for retransmission. In
the NC scheme, each node linearly combines its $M$ packets before taking turns to transmit the ensuing
mixtures. The coefficients can be generated on the fly and attached to the data payload, or tabulated a
priori. Assume the field size is large enough such that accepted coded packets are independent from each
other with very high probability. Since coded packets represent degrees of freedom (dof) rather than
distinct data packets, each node can transmit more than the required number of dof to compensate for
packet losses. Figure~\ref{fig:scheme}(b) illustrates one round of transmission. $M_k$ represents the
number of coded packets for (re)transmission. $M_k$ is a function of $p_k$, the erasure probability, and
$r_k$, the remaining number of dof needed at the BS to decode successfully. Note $M_k \geq r_k$. Since
all nodes within the network need to wakeup from sleep modes to listen to the ack, reducing the number
of ack packets effectively reduces the total energy consumption. $M_k$ should also be kept small to
minimize redundant transmissions. We want to show that an optimal number, $M_k$, $1\leq
k\leq K$, of coded packets exists to minimize the mean completion energy.

\vspace{-4pt}
\subsection{Energy Consumption}
\vspace{-2pt}

We assume sensor nodes operate in two modes: transceive and sleep. Denote the processing and
transmission energy per data packet by $E_{p, CARQ}$ and $E_{p,NC}$ for the uncoded and coded cases
respectively; also denote the reception and processing energy per ack packet by $E_{a,CARQ}$ and
$E_{a,NC}$. When in sleep mode, most circuit components are assumed to be turned off such that energy
consumption is negligible. Let $E_{a, CARQ} = \alpha {E_{p, CARQ}}$, $E_{a, NC} = \alpha {E_{p, NC}}$,
where the parameter $\alpha$ can take on different positive values depending on the circuit and protocol
designs. For example, in narrow-band systems where transmission power is approximately the same as
receiving power, $\alpha$ is the ratio between lengths of ack and data packets. For short range
ultra-wide band systems where transmission energy per bit is much smaller than the reception energy per
bit, $\alpha$ can take on large values in the range of tens to the hundreds \cite{ryckaert2005ultra, mercier2009energy,
daly2010pulsed}. Moreover, assume $E_{p,NC} = (1+\beta)E_{p,CARQ}$, where the non-negative factor
$\beta$ represents the additional energy needed to perform network coding. In later sections, we will
study the effect $\alpha$ and $\beta$ have on overall completion energy.

\vspace{-4pt}

\section{Expected Energy for Completing Transmission}\label{sec:analysis}
\vspace{-2pt}

To study the expected completion energy of uploading data from sensor nodes to the BS in a WBAN, we
model the communication process using a Markov chain. Let state $I =(i_1,\ldots, i_K)$ represent the dof
requested by the BS from nodes $(N_1,\ldots,N_K)$ for the next round of transmission, where $0 \leq i_k
\leq M$, $1\leq k \leq K$. The overall communication process initializes in state $\mathbf{M} =
(M,\ldots,M)$ and terminates in state $\mathbf{0} = (0,\ldots,0)$. Assume packet losses occur
independently across nodes, the transition probability from state $I=(i_1,\ldots,i_K)$ to state
$J=(j_1,\ldots,j_K)$ is $P_{IJ}= P_{(i_1,\ldots,i_K)(j_1,\ldots,j_K)} = \prod_{l=1}^{K}{P_{(i_k,j_k)}}$,
where $P_{(i_k,j_k)}=P_{i_kj_k}$ and
\begin{equation}
P_{ij} =\left\{
            \begin{array}{cc}
              \left(\begin{array}{c} c_{i,k} \\ i-j \\ \end{array}
                           \right)(1-p_k)^{i-j}p_k^{c_{i,k}-i+j} & \quad 0<j\leq i \\[12pt]
              \sum_{l=i}^{c_{i,k}}{\left(\begin{array}{c} c_{i,k} \\ l \\ \end{array}
                           \right)(1-p_k)^lp_k^{c_{i,k}-l}} & \quad j=0 \\
            \end{array}
          \right.
\,.\label{eq:pij}\end{equation}

\begin{figure}[!tb]
\centering
\includegraphics[width=0.65\textwidth,keepaspectratio,clip=true]{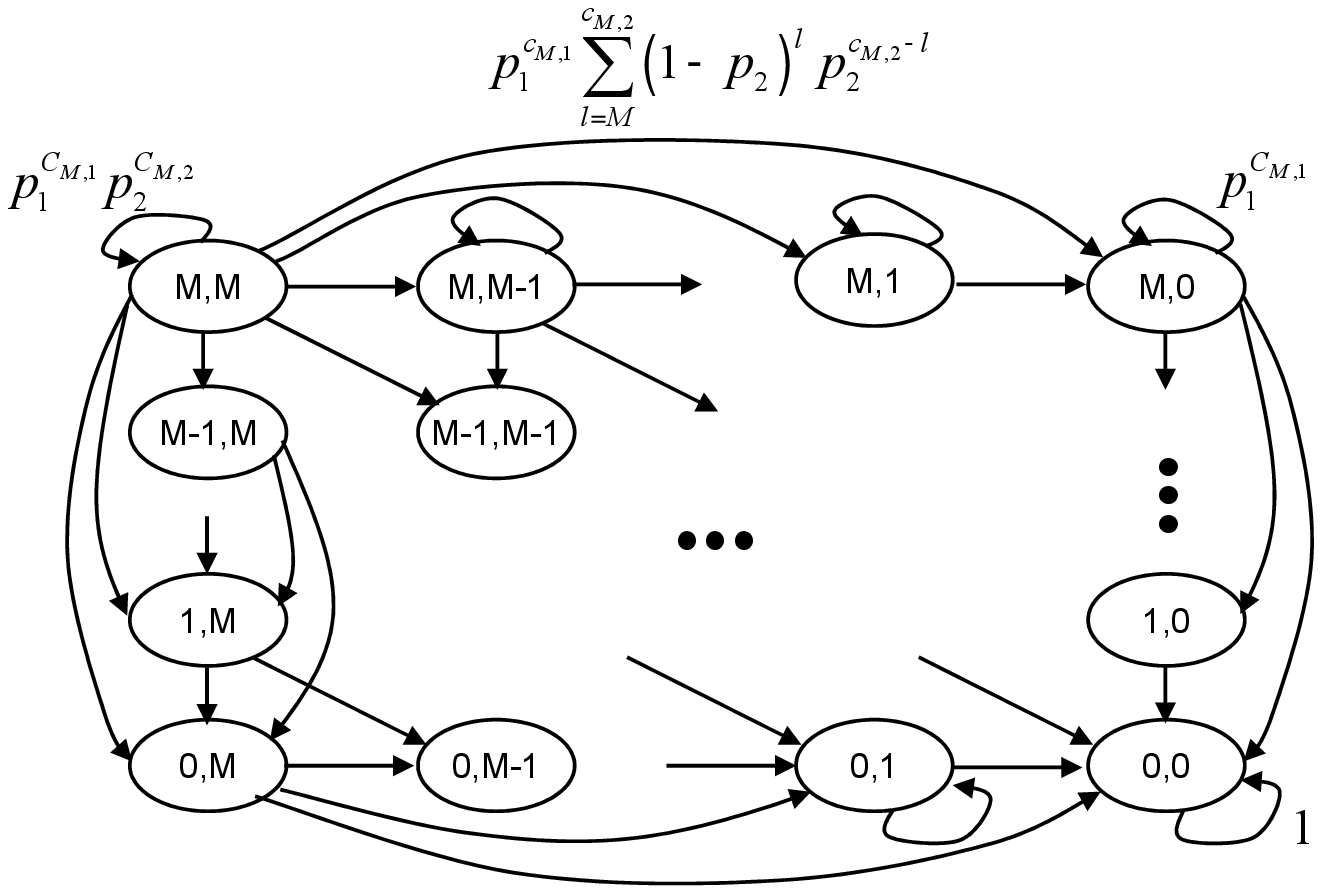}
\vspace{-6pt} \caption{Markov chain representation of the network coded scheme, number of nodes is
$K=2$.}\label{fig:mc}
\end{figure}

\noindent $c_{i,k} = M_k(p_k,i)$ denotes the number of coded packets node $N_k$ transmits when it sees a
packet erasure probability of $p_k$ and the BS requires $i$ additional dof for decoding; $c_{0,k} =0$.
The value of $c_{i,k}$ is computed by the BS. This Markov chain has $(M+1)^{K}-1$ transient states and
one recurrent state, $\mathbf{0}$, which signals completion of the transmission. Figure~\ref{fig:mc}
illustrates the case where there are $K=2$ sensor nodes within the WBAN.

Let $E_{I}$ denote the expected system completion energy when nodes $(N_1,\ldots,N_K)$ have
$(i_1,\ldots, i_K)$ dof to upload to the BS respectively, then $E_{I}$ is the expected absorption time
of this Markov chain. Let $\mathcal{Q} = \{0,\ldots, i_1\} \times \ldots \times \{0,\ldots, i_K\}$.
The following recursion holds $E_{I} = \frac{1}{1-\prod_{k=1}^{K}p_k^{c_{i_k,k}}} \left\{
E_p\sum_{k=1}^{K}c_{i_k,k}+E_{a}K + \sum_{J\in \mathcal{Q}\setminus I}{P_{IJ} E_{J}}\right\}$.
Unlike the erasure probabilities $P_{IJ}$, this expected completion energy can not be separated into
node-specific energy terms.
To minimize the expected completion energy $E_{\mathbf{M}}$, let $C =\{c_{i,k} | 1\leq i \leq M, 1\leq
k\leq K \}$, $c_{0,k} = 0, 1\leq k \leq K$, we then have $C^\ast =
\underset{C}{\operatorname{argmin}}\,{E_{\mathbf{M}}}$, $E^\ast_{\mathbf{M}} = \min_{C}E_{\mathbf{M}}$,
and the following recursion,  where $\mathbb{Z}_{M+1}=\{0,\ldots,M\}$.
\begin{align} E^\ast_{\mathbf{M}} = \min_{c_{M,1}\,\ldots, c_{M,K}}{\frac{1}{1-\prod_{k=1}^{K}p_k^{c_{M,k}}} \left\{
E_p\sum_{k=1}^{K}c_{M,k}+E_{a}K + \sum_{J\in
                                     (\mathbb{Z}_{M+1})^K
                                     \backslash \mathbf{M}}
                              P_{\mathbf{M}J} E^\ast_{J} \right\}}
\,.\end{align}

One approach to this optimization is to ignore the integer constraints, and solve for $c_{i,k}$
iteratively by finding values that set the partial derivatives of the objective function to zero.
However, it can be shown that no closed-form solution exists. Also since there are $(M+1)^K$ states in
the Markov chain. As $M$ and $K$ increase, the computational complexity becomes prohibitive for a
practical system. An alternative is to perform exhaustive numerical searches for the optimal values
$C^\ast$ on an integer grid. For given values of $\{p_k|1\leq k \leq K\}$, $E_p$, and $E_a$, we can
recursively search on an $M$ dimensional space of non-negative integers to find $C^\ast$. We will show
numerical examples in the next section for such an optimal scheme. In a practical implementation, the
computation task is imposed on the BS, not individual sensor nodes. Neither do the results need to
be computed in real time. Instead, pre-computed values can be stored in a look-up table according to
different packet erasure probabilities. The exact quantization required to balance the accuracy and
required memory is a topic for future studies.

\vspace{-4pt}

\section{Numerical Examples}\label{sec:simulations}
\vspace{-2pt}

In this section, we provide numerical examples for the CARQ and NC schemes to study the amount of energy reduction offered by network coding as system parameters vary.

Table~\ref{tbl:egSolution} lists explicitly the solution to the optimization problem stated in
Section~\ref{sec:analysis} when there are $K=2$ nodes within the network, each having $M=4$ data packets
to upload, $p_1 = 0.2$, $p_2 = 0.4$. Assume $\alpha = 1$ and $\beta = 0$.  The first column(row) states
the remaining number of dof required by the BS from node $N_1$($N_2$). Transmission initiates at $(i_1,
i_2) = (4,4)$, and terminates at $(0,0)$. Since $N_2$ sees a more challenging channel, it sends more
coded packets than $N_1$, when the same number of dof is requested. Observe that the number of coded packets
sent by $N_2$ actually increases from $6$ to $7$ when $i_2 = 4$, and $i_1$ is decremented from $4$ to $0$.
This is because $N_1$ is expected to complete its data transmission in a small number of rounds, thus
$N_2$ would want to send more data packets such that it also completes its data transmission in a small
number of rounds, to reduce the total number of times both wake up to listen to ack signals. The optimal
solution minimizes the sum of all energy terms, taking into account of future rounds of retransmissions,
and tries to reduce possible energy wastes. The optimal expected total completion energy is found to be $16.46E$,
larger than $14E$ shown in Table~\ref{tbl:example}, which illustrates only one possible channel realization.

\begin{table}[t!]
\centering \caption{Optimal numbers of coded packets to transmit by each node when dof requested by the base
station is $i_1$ from node $N_1$ and $ i_2$ from node $N_2$; $M = 4$, $K=2$, $p_1 = 0.2$, $p_2 = 0.4$,
$\alpha = 1$, $\beta = 0$.} \label{tbl:egSolution}\vspace{-8pt}
\begin{tabular}{|c|c|c|c|c|c|c|}
\hline
$i_1\backslash i_2$ & $ 4$ & $ 3$ & $2$ & $1$ & $ 0$ \\\hline %
$ 4$ & $(5,6)$ & $(5,5)$ & $(5,3)$ & $(5,2)$ & $(5,0)$ \\\hline %
$ 3$ & $(3,6)$ & $(4,5)$ & $(4,3)$ & $(4,2)$ & $(4,0)$ \\\hline %
$ 2$ & $(2,6)$ & $(2,5)$ & $(2,3)$ & $(2,1)$ & $(3,0)$ \\\hline %
$ 1$ & $(1,6)$ & $(1,5)$ & $(1,3)$ & $(1,2)$ & $(1,0)$ \\\hline %
$ 0$ & $(0,7)$ & $(0,5)$ & $(0,3)$ & $(0,2)$ & $(0,0)$ \\ %
\hline
\end{tabular}\vspace{-5pt}
\end{table}

\begin{figure}[!tb]
\begin{minipage}{.46\linewidth}
  \centering
  \centerline{\epsfig{figure=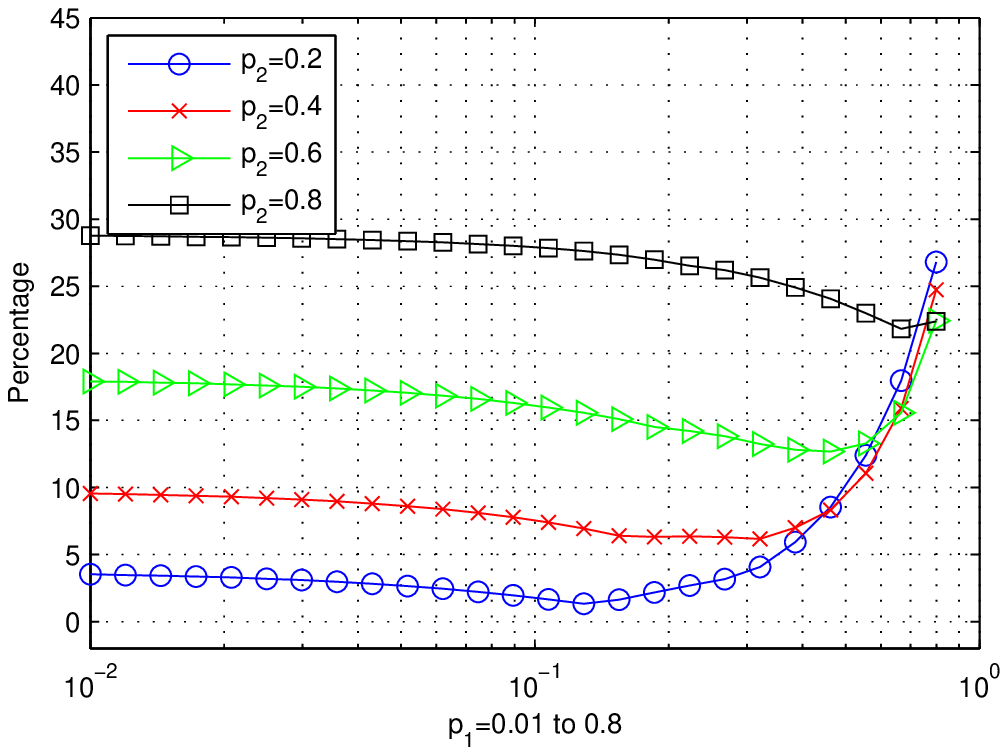, width = 6.2cm}}
  \vspace{-6pt}
  \caption{Percentage reduction in expected completion energy per
accepted data packet as $p_2$ varies; $M = 4$, $K = 2$, $\alpha = 1$, $\beta =
0$.}\label{fig:summary_p_alpha1}
\end{minipage}
\hspace{.06\linewidth}
\begin{minipage}{.46\linewidth}
  \centering
  \centerline{\epsfig{figure=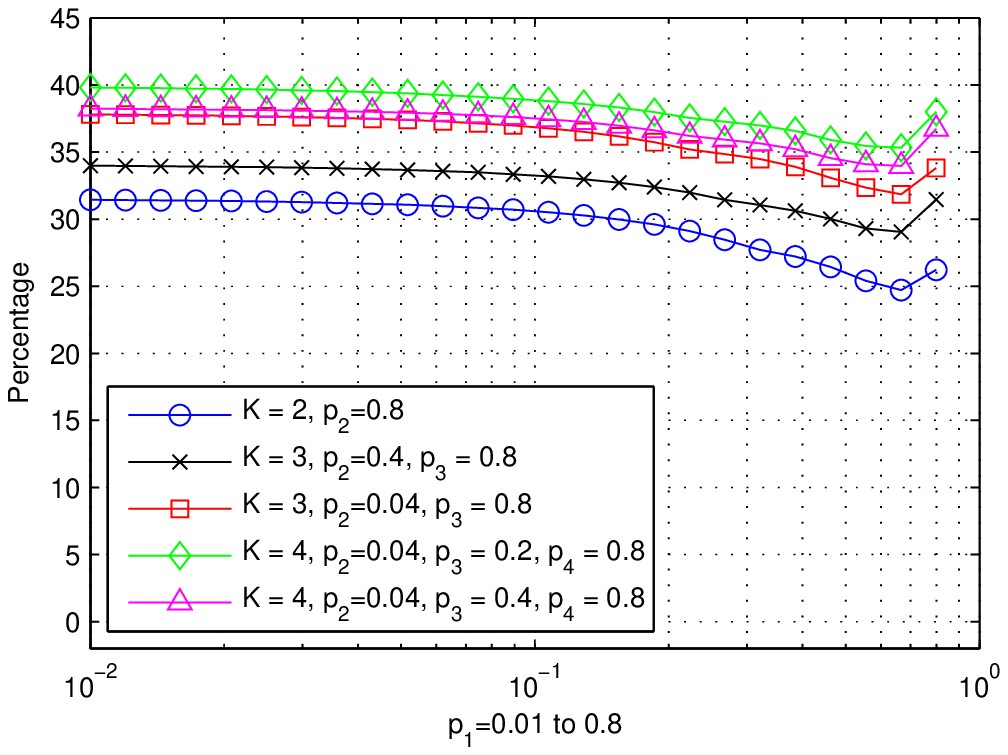, width = 6.2cm}}
  \vspace{-6pt}
  \caption{Percentage reduction in expected completion energy as $K$ is varied, $M = 2$,
  $\alpha = 1$, $\beta = 0$.}\label{fig:summary_K_alpha1}
\end{minipage}\vspace{-12pt}
\end{figure}

Figure~\ref{fig:summary_p_alpha1} extends the example in Table~\ref{tbl:egSolution} to summarize the
percentage reduction in expected completion energy per accepted data packet achieved by the NC scheme
when packet erasure probabilities vary. A packet is said to be accepted by the BS if it is received
successfully, and the percentage reduction is computed with respect to the CARQ scheme. Again, we assume
$\alpha = 1$, i.e., $E_{p, NC} = E_{a, NC}$, $E_{p, CARQ} = E_{a, CARQ}$, and $\beta = 0$, i.e., $E_{p,
NC} = E_{p,CARQ}$. The horizontal axis represents variations in the packet erasure probability of node
$N_1$. When the erasure probability of $N_2$ increases from 0.2 to 0.8, the amount of reduction in
expected completion energy per accepted data packet increases from $3.5\%$ to about $29\%$. Although not
shown explicitly in this graph, the energy gain is derived from reduced number of transmission rounds.
As $p_2$ degrades, the actual amount of energy spent for each accepted data packet increases, because
more retransmissions are expected. Since depletion occurs more quickly when the channel condition
worsens, the increased amount of saving is beneficial in extending the lifetime of sensor nodes. Another
observation from Figure~\ref{fig:summary_p_alpha1} is that the curves take a dip at different values of
$p_1$. The locations of these minima correspond approximately to the values of $p_2$ in each case. This
is because the NC scheme achieves higher energy reduction when nodes experience more
asymmetric channel conditions. When packet losses occur asymmetrically, nodes with more reliable
channels complete data transmission quickly; yet they are forced to wake up for the ack signal
repeatedly until other nodes with less reliable channels complete their transmissions. When nodes see
similar channel conditions, with high probability, all nodes have non-zero number of
packets to send each round, hence not as much energy is wasted in listening to the ack signals.

Figure~\ref{fig:summary_K_alpha1} considers the more general case when the number of sensors within the
network is increased, and $M$ is set to 2 for simplicity.
More gains can be achieved
when more nodes are included, especially under asymmetric channel conditions. Although not shown
explicitly here, we also compared numerical results under different network coding parameters. When the
generation size $M$ is varied, the amount of energy gain over CARQ is higher if $M$ is smaller, with a
decrease of approximately $10\%$ as $M$ increases from $2$ to $10$. This is because reception energy is
amortized over more data packets.  The NC scheme can also be shown to improve the throughput of the
system.

\begin{figure}[!tb]
\begin{minipage}{.46\linewidth}
  \centering
  \centerline{\epsfig{figure=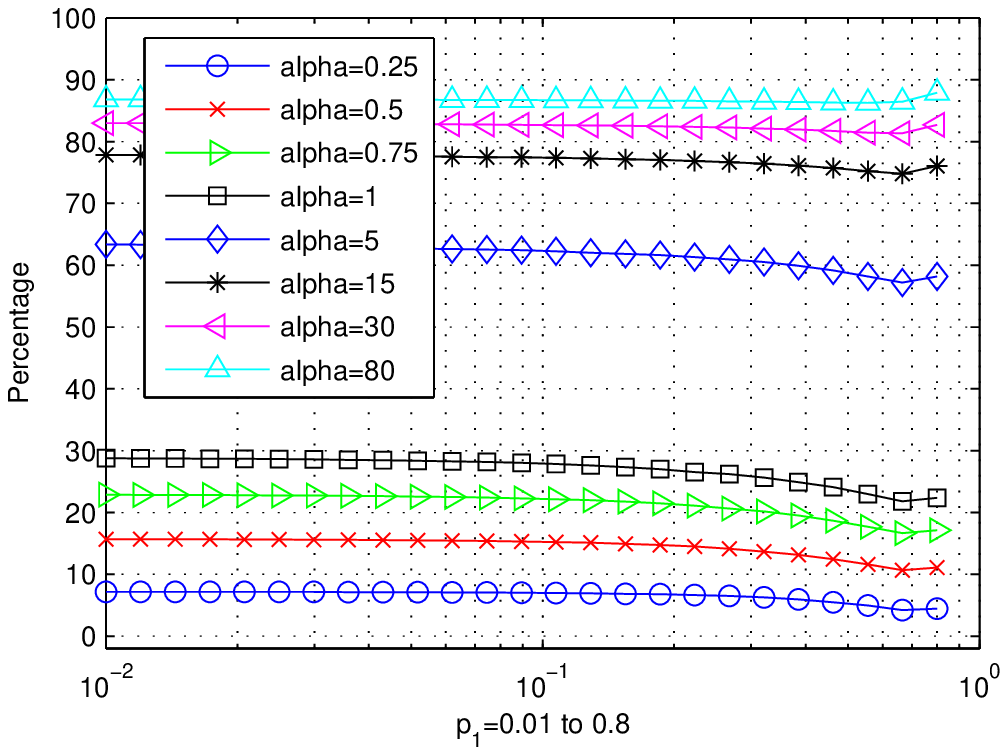, width = 6.2cm}}
  \vspace{-6pt}
 \caption{Percentage reduction in expected completion energy when $\alpha$ is varied, $M = 4$,
$K = 2$, $\beta = 0$, $p_2 = 0.8$.}\label{fig:summary_alpha_M4_p0p8}
\end{minipage}
\hspace{.06\linewidth}
\begin{minipage}{.46\linewidth}
  \centering
  \centerline{\epsfig{figure=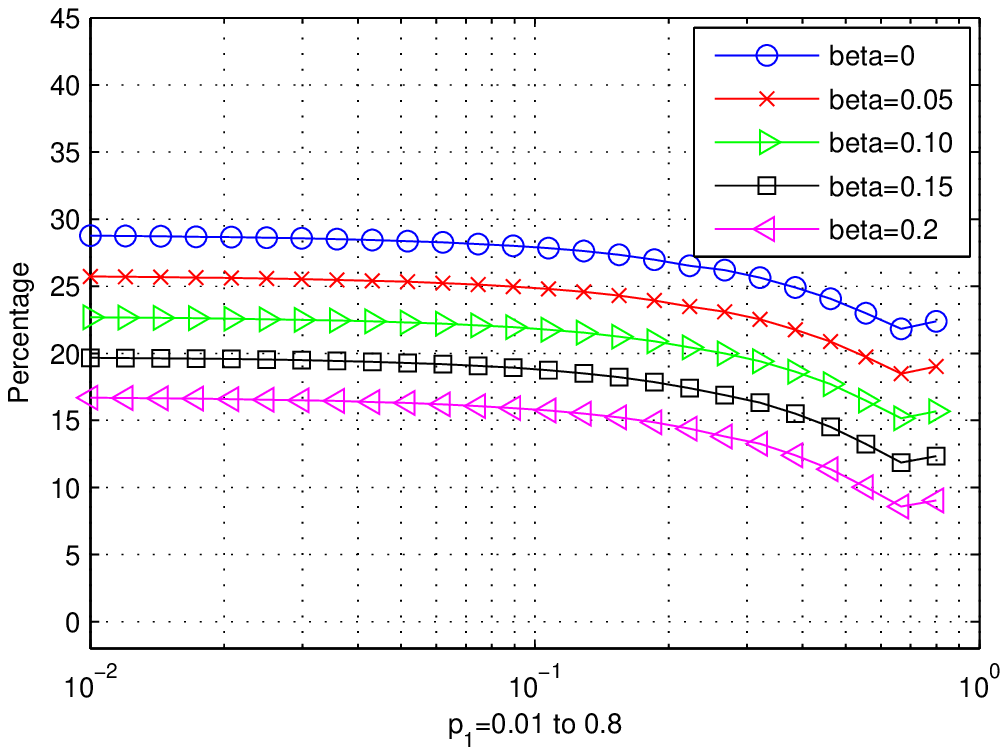, width = 6.2cm}}
  \vspace{-6pt}
\caption{Percentage reduction in expected completion energy when $\beta$ is varied, $M = 4$, $K = 2$,
$\alpha = 1$, $p_2 = 0.8$.}\label{fig:summary_beta_M4_p0p8_alpha1}
\end{minipage}\vspace{-12pt}
\end{figure}

So far we have examined reduction in completion energy achievable through NC when the reception energy
$E_a$ per ack packet is the same as the transmission energy $E_p$ per data packet, i.e., $\alpha = 1$.
The actual value of $\alpha$ is dependent on the circuit architectures of the transmitter and receiver,
and the data and ack packet payload design. For example, $\alpha$ can be on the order of 1 for a narrow
band system, but can be one or two orders of magnitude larger for an ultra-wide band system
\cite{ryckaert2005ultra,mercier2009energy,daly2010pulsed}. Figure~\ref{fig:summary_alpha_M4_p0p8} shows the reduction in
completion energy in using NC over CARQ when $\alpha$ is varied, where $E_a = \alpha E_p$, $E_p=E_{p,NC}
= E_{p,CARQ}$ and $E_a = E_{a,NC} = E_{a,CARQ}$. Again, the computation is conducted over a $2$-node
network for simplicity. The observed gain is not as significant as $29\%$ when $\alpha$ decreases from
$1$, for data transmission energy much outweighs ack reception energy. However, as $\alpha$ increases to
values higher than $15$, we can achieve up to $87\%$ in energy reduction, i.e., 5 times less energy.
This is equivalent to extending the lifetime of a sensor node by a factor of 5.

Another assumption we have made explicitly in previous examples is that the average transmission energy
$E_p$ is the same for both NC and CARQ. In an actual implementation, the NC scheme may require
non-negligible energy overheads for coding. Figure~\ref{fig:summary_beta_M4_p0p8_alpha1} compares the
completion energy of the two schemes when $E_{p,NC} = (1+\beta)E_{p,CARQ} = (1+\beta) \alpha
E_{a,CARQ}$, $\alpha = 1$, and $E_{a,NC} = E_{a,CARQ}$. Here the energy advantage is lessened because of
the added cost of coding. Nonetheless, even with a $20\%$ overhead in coding, we can still achieve an
energy reduction of about $17\%$. 

\vspace{-4pt}

\section{Conclusion}\label{sec:conclusions}
\vspace{-2pt}

We proposed a network coded scheme to help improve energy efficiency of wireless body area networks.
Assuming that the different channel conditions experienced by individual nodes are known at the base
station, the base station can request each sensor node to send an optimal number of coded packets,
taking into account anticipated packet losses during transmission, and energy needed for receiving
control signals. We show with numerical examples that in a two-node star, when transmitting a data
packet and receiving an ack packet cost approximately the same amount of energy, the network coded
scheme can achieve up to $29\%$ percent reduction in expected completion energy per accepted data packet
compared to the CARQ scheme. When receiving costs a lot more than transmitting, network coding can
reduce energy use by up to $87\%$. We also shown with numerical examples that the amount of energy gain
achievable through coding increases as more nodes are added to the network, and when nodes see more
asymmetric channel conditions.

\vspace{-4pt}

\section{Acknowledgment}\label{sec:acknolwdgement}
This material is based upon work supported under subcontract \# RA306-S1
issued by the Georgia Institute of Technology, by the Claude E. Shannon Research Assistantship awarded by the Research Laboratory of Electronics, MIT, and by the NSERC Postgraduate Scholarship (PGS) issued by the Natural
Sciences and Engineering Research Council of Canada. This work is also partially funded by the European
Commission under grant FP7-INFSO-ICT-215252 (N-Crave Project).

\bibliographystyle{splncs03}
\bibliography{references}
\end{document}